\documentclass[12pt]{iopart}
\usepackage{graphicx}
\usepackage{cite}

\begin{document}

\title[A new empirical formula for $\alpha$-decay half-life and decay chains of Z$=$120 isotopes]{A new empirical formula for $\alpha$-decay half-life and decay chains of Z$=$120 isotopes}

\author{G. Saxena$^1$, A. Jain$^{1,2}$, and P. K. Sharma$^3$}

\address{$^1$Department of Physics (H\&S), Govt. Women Engineering College, Ajmer-305002, India }
\address{$^2$Department of Physics, School of Basic Sciences, Manipal University Jaipur, Jaipur-303007, India}
\address{$^3$Govt. Polytechnic College, Rajsamand-313324, India }
\ead{gauravphy@gmail.com}

\vspace{10pt}
\begin{indented}
\item[]July 2021
\end{indented}

\begin{abstract}
Experimental $\alpha$-decay half-life, spin, and parity of 398 nuclei in the range 50$\leq$Z$\leq$118 are utilized to propose a new formula (QF) with only 4 coefficients as well as to modify the Tagepera-Nurmia formula with just 3 coefficients (MTNF) by employing nonlinear regressions. These formulas, based on reduced mass ($\mu$) and angular momentum taken away by the $\alpha$-particle, are ascertained very effective for both favoured and unfavoured $\alpha$-decay in addition to their excellent match with all (Z, N) combinations of experimental $\alpha$-decay half-lives. After comparing with similar other empirical formulas of $\alpha$-decay half-life, QF and MTNF formulas are purported with accuracy, minimum uncertainty and deviation, dependency on least number of fitted coefficients together with less sensitivity to the uncertainties of $Q$-values. The QF formula is applied to predict $\alpha$-decay half-lives for 724 favoured and 635 unfavoured transitions having experimentally known $Q$-values. Moreover, these available $Q$-values are also employed to test various theoretical approaches viz. RMF, FRDM, WS4, RCHB, etc. along with machine learning method XGBoost for determining theoretical $Q$-values, incisively. Thereafter, using $Q$-values from the most precise theoretical treatment mentioned above along with the proposed formulas, probable $\alpha$-decay chains for Z$=$120 isotopes are identified.
\end{abstract}
\noindent{\it keywords}:  Superheavy nuclei; $\alpha$-decay half-life; Empirical formulas; Favoured and unfavoured $\alpha$-decay.
%\maketitle

\section{Introduction}

$\alpha$-decay, one of the dominant mode of decay in heavy and superheavy nuclei, has its history from the quantum mechanical formulation by Gamow \cite{gamow1928} accompanied by Gurney and Condon \cite{condon1928} in 1928. Ever since $\alpha$-decay is used as a probe to explore nuclear structural properties including identification of new elements of periodic chart \cite{hofmann2000,hamilton2013,ogan2015,heenen2015,oganrpp2015,oganpt2015,hofmann2016,dull2018,nazar2018,giuliani2019}. The synthesis of superheavy elements through cold-fusion and hot-fusion reactions in the laboratories has prevailed remarkable success which is complemented by observation of decay chains of $^{293}$117
and $^{294}$117 with 11 new nuclei \cite{Oganessian2010} as well as decay chains of the heaviest element $^{294}$118 \cite{ogan2006}. In addition, eleven new heaviest isotopes of elements Z$=$105 to Z$=$117 are identified affirming the concept of the island of enhanced stability (near N$\sim$184) for superheavy nuclei (SHN) \cite{ogan2011} which is further strengthened by various properties of the decay of the heaviest nuclei having Z$=$112–118 \cite{Ogannpa2015}. Already, several attempts have aimed to synthesize heavier elements with Z$\geq$118 and in near future, new and more precise experimental facilities are expected in search of new SHN \cite{hofmann2016,Oganessian2009}. Recently, a new decay chain of $^{294}$Og along with prospects for reaching new isotopes $^{295,296}$Og \cite{brewer2018} has been observed. Likewise, reactions of formation of $^{294}$Ts and $^{294}$Og nuclides has been analyzed to capitalize the future possibilities of synthesizing new elements with Z$=$119 and Z$=$120 \cite{voinov2020} by measuring $\alpha$-decay half-life.\par

Theoretically, one way to determine the $\alpha$-decay half-life is based upon empirical/semiempirical formulas \cite{geiger1911,vss1966,brown1992,royer2000,horoi2004,renA2004,renB2008,qi2009,poenaru2006,poenaru2011,manjunatha2019}: dependent mainly on $\alpha$-decay energy ($Q$-value). Many of these formulas are modified time to time in variety of forms \cite{sobi1989,parkho2005,budaca2016,royer2010,royer2020,poenaru2007,akrawymrf2018,akrawyijmpe2018,akrawyprc2019,newrenA2019} to match with latest experimental fits along with our very recent works \cite{singh2020,saxenajpg2020}. Besides the dependencies of $\alpha$-decay accounted in various formulas mentioned above, the centrifugal potential distinctly contributes in $\alpha$-transitions \cite{denisov2009} which necessitates the intromission of spins and parities of the parent and daughter nuclei together with the angular momentum of the emitted $\alpha$-particle for accurate consideration of the $\alpha$-transitions \cite{bohr1975}. Based on selection rules \cite{denisov2009} for the minimal orbital angular momentum ($l_{min}$) of the emitted $\alpha$-particle, $\alpha$-transitions are categorized into favoured ($l_{min}$$=$0) and unfavoured ($l_{min}$$\neq$0) transitions. Few contemplation with the use of centrifugal term are already pointed in Refs. \cite{royer2010,denisov2009prc,dong2010npa,qian2012,wang2015}, accompanied by several modified versions \cite{royer2020,akrawymrf2018,akrawyijmpe2018,akrawyprc2019,newrenA2019,akrawy2018} which have investigated favoured and unfavoured $\alpha$-transitions demonstrating crucial role of orbital angular momentum ($l_{min}$) of the emitted $\alpha$-particle for describing $\alpha$-decay half-life, precisely.\par

In view of the above recital, there are a variety of formulas evaluating $\alpha$-decay half-life in the superheavy realm but still, preciseness of a formula for the entire range of periodic chart without imposing the use of lots of coefficients is an essential requirement for the prediction of unknown territory of superheavy nuclei. This possibility has invoked us to propose a new empirical formula which can be applied for the entire domain of $\alpha$-decay with the no. of fitted coefficients as less as possible and substantially accurate to calculate $\alpha$-decay half-life. Additionally, the formula should be embedded with less uncertainty or deviation, and the less sensitive to the uncertainties usually found in $Q$-values. With this focus, we propose our new formula (QF formula) followed by a modified version of Tagepera-Nurmia formula (MTNF). These formulas are fitted using $scipy.optimize.curve\_fit$ library \cite{scipy} along with Levenberg-Marquardt algorithm which is used for non-linear least-squares fit \cite{wood}. We first compare results of our formulas with already established similar formulas as well as with the half-lives of experimentally known decay chains. Thereafter, we determine half-lives of probable decay chains for Z$=$120 isotopes consisting neutron numbers in the range 168$\leq$N$\leq$184.\par
\section{Formalism and Calculations}
To set up a new formula for determining $\alpha$-decay half-life, we have applied different fitting procedures including linear and nonlinear regressions. Since, the non-linear regression determines parameters that minimize the error by calculating the squares distances of the data points with the model \cite{motulsky1987,krakovska2019}. Therefore, as a result, we have fitted the second-order polynomial term of Z$_{d}^{m}$/$\sqrt{Q}$ including dependencies on reduced mass ($\mu$) and centrifugal potential. The higher-order polynomials are neglected to avoid overfitting of the data. This new formula as well as the power of Z$_d$, are fitted by using 398 $\alpha$-decay half-lives in the range 50$\leq$Z$\leq$118. The new formula (QF) having linear and quadratic term of Z$_{d}^{0.6}$/$\sqrt{Q}$ is given by:
 \begin{equation}\label{QF}
log_{10}T_{1/2}^{\alpha}(sec.) = a\sqrt{\mu}\left(\frac{Z_d^{0.6}}{\sqrt{Q}}\right)^2 + b\sqrt{\mu}\left(\frac{Z_d^{0.6}}{\sqrt{Q}}\right)+c+dl(l+1),
\end{equation}
where Z$_{d}$ is proton number of daughter nucleus after $\alpha$-emission and $Q$ is disintegration energy in MeV. The coefficients (a,b,c,d) obtain by the fitting are given in Table \ref{coeff}.\par
 In 1961, Tagepera and Nurmia \cite{tnf1961} have formulated the semi-empirical formula for the logarithmic half-lives (in years) of $\alpha$-decay. In the present work, this formula is modified by including reduced mass and centrifugal potential term, which shapes as:\\
\begin{equation}\label{mtnf}
log_{10}T_{1/2}^{\alpha}(sec.) = a\sqrt{\mu}(Z_d Q^{-1/2}-Z_{d}^{2/3})+ b+cl(l+1).
\end{equation}
This modified Tagepera and Nurmia formula (MTNF) is fitted for the new experimental data \cite{nndc} using scipy \cite{saxenajpg2020,scipy} and the coefficients are readjusted accordingly which are given in Table \ref{coeff}.

In both formulas (Eqns. \ref{QF} and \ref{mtnf}), the last term represents contribution of centrifugal potential where $l$ is the minimum angular momentum taken away by $\alpha$-particle, which can be obtained by following selection rules, based on conservation of angular momentum and parity \cite{denisov2009}:
\begin{equation}
   l_{min}=\left\{
    \begin{array}{ll}
       \triangle_j\,\,\,\,\,\,
       &\mbox{for even}\,\,\triangle_j\,\,\mbox{and}\,\,\pi_{p} = \pi_{d}\\
       \triangle_{j}+1\,\,\,\,\,\,
       &\mbox{for even}\,\,\triangle_j\,\,\mbox{and}\,\,\pi_{p} \neq \pi_{d}\\
       \triangle_{j}\,\,\,\,\,\,
       &\mbox{for odd}\,\,\triangle_j\,\,\mbox{and}\,\,\pi_{p} \neq \pi_{d}\\
       \triangle_{j}+1\,\,\,\,\,\,
       &\mbox{for odd}\,\,\triangle_j\,\,\mbox{and}\,\,\pi_{p} = \pi_{d}\\
      \end{array}\right.
      \label{lmin}
\end{equation}
where $\triangle_j$ = $|j_p - j_d|$ with j$_{p}$, $\pi_{p}$, j$_{d}$, $\pi_{d}$, being the spin and parity values of the parent and daughter nuclei, respectively. For the purpose of fitting, parity and spin of 398 nuclei are taken from the latest evaluated nuclear properties table NUBASE2020 \cite{audi2020}.\par

It is appropriate to mention here that the above-mentioned formulas are proposed after probing their merit over a few other fittings viz. including (i) asymmetry terms ($I + I^2$, where $I$$=$$\frac{N-Z}{A}$ is the isospin) only, and (ii) asymmetry terms as well as the centrifugal potential term. It is found that our proposed formulas without adding too many terms and coefficients result effectively well, as attempted for favoured and unfavoured $\alpha$-decay by Deng \textit{et al.} \cite{royer2020}, and therefore, expeditiously qualify to predict $\alpha$-decay half-lives in the unknown superheavy region of experimental interest (Z$>$118). \par
Since, in this part of periodic chart the $\alpha$-decay competes with spontaneous fission: the decay which breaks the chain of $\alpha$-decay. Therefore, to estimate probable $\alpha$-decay chains the half-lives of $\alpha$-decay should be compared with the half-lives of spontaneous fission (SF), for which we use modified version of Bao formula \cite{bao2015} from our recent work \cite{saxenajpg2020} that is given by:
\begin{table}
\caption{\label{coeff}The coefficients of QF formula and MTNF.}
\centering
\footnotesize
\begin{tabular}{@{}lcccc}
\br
Formula&a&b&c&d\\
\mr
QF & -0.3980&	9.2318&	-75.3725&	0.0354\\
MTNF & 0.7208 &	-17.7056 &	0.0281&-\\
\br
\end{tabular}\\
\end{table}
\normalsize
\begin{equation}\label{baoSF}
log_{10}T_{1/2}^{SF} (s) = c_1 + c_2 \left(\frac{Z^2}{(1-kI^2)A}\right)+ c_3 \left(\frac{Z^2}{(1-kI^2)A}\right)^2 + c_4 E_{s+p}.
\end{equation}

Here k$=$2.6 and other coefficients are c$_2$$=$$-$37.0510, c$_3$$=$0.3740, c$_4$$=$3.1105.
The values of c$_1$ for various sets of nuclei are c$_1$(e-e)$=$893.2644, c$_1$(e-o)$=$895.4154, c$_1$(o-e)$=$896.8447 and c$_1$(o-o)$=$897.0194.
%%%%%%%%%%%%%%%%%%%%%%%%%%%%%%%%%%%%%%%%%%%%%%%%%%%%%%%%%%%%%%%%%%%%%%%%%%%%%%%%%%%%%%%%%%%%%%%%%%%%%%%%%%%%%%%%%%%%%%%%%%%%%%%%%%%%%%%%%%%%%%%%%%%%%%
\section{Results and discussions}
Before all else, we enumerate standard deviation ($\sigma$), uncertainty ($u$), average deviation factor ($\overline{x}$) and mean deviation ($\overline{\delta}$) (as mentioned in the following self-explanatory equations) for several empirical/semi-empirical formulas of $\alpha$-decay half-lives. These all statistical parameters are mentioned for 8 empirical/semi-empirical formulas along with number of fitted coefficients lined in the particular formula in Table \ref{rmse-alpha}.

\begin{equation}
\sigma = \sqrt{\frac{1}{N_{nucl}-1}\sum^{N_{nucl}}_{i=1}\left(log\frac{T^i_{th}}{T^i_{exp}}\right)^2};
\end{equation}

\begin{equation}
u =  \sqrt{\frac{1}{N_{nucl}(N_{nucl}-1)}\sum^{N_{nucl}}_{i=1}\left(log\frac{T^i_{th}}{T^i_{exp}}-\mu \right)^2};
\end{equation}

%\begin{equation}
%\mu = \frac{1}{N_{nucl}}\sum^{N_{nucl}}_{i=1}\left(log\frac{T^i_{th}}{T^i_{exp}}\right)
%\end{equation}

\begin{equation}
\overline{x} = \frac{1}{N_{nucl}}\sum^{N_{nucl}}_{i=1}\left(\frac{|logT^i_{exp}-logT^i_{th}|}{logT^i_{exp}}\right);
\end{equation}

\begin{equation}
\overline{\delta} = \frac{1}{N_{nucl}}\sum^{N_{nucl}}_{i=1}\left|log\frac{T^i_{th}}{T^i_{exp}}\right|.
\end{equation}

Several latest formulas or the modified/refitted versions of few widely known formulas \cite{newrenA2019,akrawy2018,akrawymrf2018,akrawyijmpe2018,akrawyprc2019,royer2020} which also incorporate centrifugal term alike our proposed formulas (QF and MTNF) are compared. We have considered only those few formulas which are fitted recently (from the year 2018 onwards) to make the comparison justifiable in terms of used data set, and therefore, we believe that addition of few new data (if any) after 2018 will not impact overall trend of results of Table \ref{rmse-alpha} in reality.
\begin{table}
\caption{\label{rmse-alpha}Standard deviation ($\sigma$), uncertainty ($u$), average deviation factor ($\overline{x}$) and mean deviation ($\overline{\delta}$) of various formulas of $\alpha$-decay half-lives calculated for nuclei in the range of 50$\leq$Z$\leq$118. No. of fitted coefficients used in the respective formula are also mentioned.}
\centering
\footnotesize
\begin{tabular}{l@{\hskip 0.1in}c@{\hskip 0.1in}c@{\hskip 0.1in}c@{\hskip 0.1in}c@{\hskip 0.1in}c}
\hline
\hline
Formula    &No. of Coefficients                     & $\sigma$& $u$ & $\overline{x}$&$\overline{\delta}$\\
\hline
QF (Present work)                     &4     &  1.3018  & 0.0653    &  1.5901  &  0.6432     \\
MTNF (Present work)                     &3     &  1.5091  & 0.0756    &  1.4967  &  0.7679   \\
%Dong (2010)                             &12    &  1.6774  & 0.0841   &  1.6926  &  0.6116   \\
Royer (2020)                            &8     &  1.6958  & 0.0850    &  1.7260  &  0.6400   \\
MYQZR (2019)                            &24    &  1.7973  & 0.0901   &  1.9668  &  0.6998    \\
%Wang (2015)                             &17    &  1.7032  & 0.0854   &  2.2757  &  0.8476   \\
Akrawy (2018)                           &24    &  1.8305  & 0.0918   &  2.1230  &  0.8513    \\
Modified RenB (2019)                    &24    &  1.8653  & 0.0935   &  2.1275  &  0.7997    \\
%Analytical Formula of Royer (RB) (2010) &18    &  1.8725  & 0.0939   &  2.0338  &  0.8192   \\
MRF (2018)                              &28    &  1.8817  & 0.0943   &  2.1838  &  0.8895    \\
%DK1 (2009)                              &20    &  1.9957  & 0.1000   &  2.3219  &  0.9087   \\
DK2 (2018)                              &20    &  2.0640  & 0.1035   &  2.1120  &  0.8474    \\
%YQZR (2012)                             &16    &  2.9182  & 0.1463   &  5.1988  &  2.1444   \\
\hline
\hline
\end{tabular}\\
\end{table}
\normalsize
\begin{figure}[!htbp]
\centering
\includegraphics[width=0.8\textwidth]{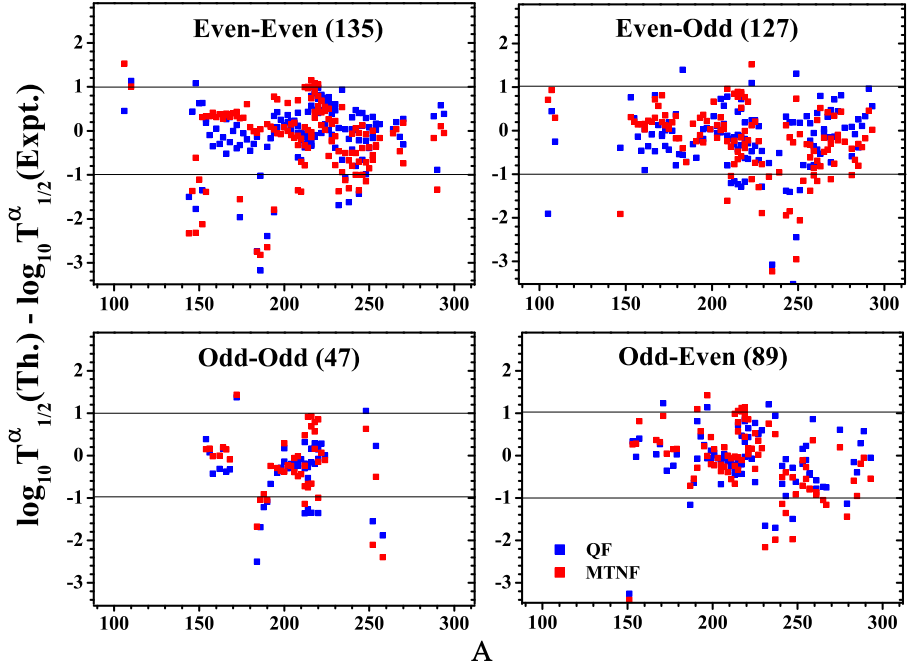}
\caption{Errors in $log_{10}T_{1/2}^{\alpha}$ values (in sec.) calculated by using QF and MTNF formulas for all sets of nuclei in the range  of 50$\leq$Z$\leq$118.}\label{fig1}
\end{figure}
The differences between experimental data and calculations from QF and MTNF formulas of the logarithmic values of $\alpha$-decay half-life with respect to mass number A are shown in various panels of Fig. \ref{fig1} for 135 even-even, 127 even-odd, 47 odd-odd, and 89 odd-even nuclei in the range of 50$\leq$Z$\leq$118. It is gratifying to note from Table \ref{rmse-alpha} and from Fig. \ref{fig1} that newly proposed QF formula, as well as modified TNF formula result with lowest uncertainty and minimum error and hence, accomplish to predict $\alpha$-decay half-lives of various nuclei throughout the periodic chart. By and large, our proposed formulas are found quite accurate, however, there are few data points in Fig. \ref{fig1} at which the proposed formulas are off by few orders of magnitude difference. The applicability of the aforementioned formulas is justified with the fact that the same set of coefficients (4 for QF and 3 for MTNF) is capable to bring out the $\alpha$-decay half-lives for all combinations of Z and N (even-even, even-odd, odd-even, and odd-odd) in an explicit manner. This specific character meliorates QF and MTNF formulas as compared to several other formulas listed in Table \ref{rmse-alpha}. In the Table \ref{blocking}, we have listed standard deviation of errors for all even-odd, odd-even, and odd-odd sets of nuclei using QF and MTNF formulas and compared with the latest fitted formula (Royer 2020) \cite{royer2020} in which blocking effect of the unpaired nucleon \cite{xu2005,sun2017} has been incorporated. It is indulging to note from Table \ref{blocking} that our proposed formulas are capable to produce reasonably precise half-lives of odd A and odd-odd nuclei.\par

\begin{table}
\caption{\label{blocking}Standard deviation ($\sigma$) in QF and MTNF formulas for even-odd, odd-even and odd-odd nuclei. These values are compared with latest similar fitted formula (Royer 2020) \cite{royer2020}.}
\centering
\footnotesize
\begin{tabular}{l@{\hskip 0.4in}c@{\hskip 0.4in}c@{\hskip 0.4in}c}
\hline
\hline
Sets of Nuclei    & QF&MTNF&Royer 2020 \\
\hline
Even-Odd & 1.1543 & 1.4247 & 1.6957 \\
Odd-Even & 1.8287 & 1.8133 & 1.8732 \\
Odd-Odd  & 1.0040 & 0.9566 & 1.0734 \\
\hline
\hline
\end{tabular}\\
\end{table}

Another attribution of our proposed formulas is the sensitivity of half-life on uncertainties of $Q$-values. Lesser the sensitivity on uncertainties of $Q$-value leads more the accuracy in determining half-life in the unknown territory where exact $Q$-values are not available. To check such sensitivity of our proposed formula, we have calculated ROC (rate of change) \cite{singh2020} of half-life with respect to $Q$-value using the following equation:
      \begin{equation}
        ROC= \left[\frac{T_Q}{T_{Q + 0.05 MeV}}-1\right]\times100\%,
      \end{equation}

\begin{table}[!htbp]
\caption{\label{roc}ROC of $\alpha$-decay half-life with respect to the uncertainties of $Q$-value for considered empirical formulas.}
\centering
\begin{tabular}{l@{\hskip 0.7in}c}
\hline
\hline
Formula&ROC\\
\hline
MTNF (Present work)                           & 53.16     \\
QF (Present work)                           & 55.30     \\
%Dong (2010)                                   & 64.61    \\
%Wang (2015)                                   & 65.10    \\
Royer (2020)                                  & 65.25     \\
%YQZR (2012)                                   & 66.61    \\
Akrawy (2018)                                 & 67.09    \\
MRF (2018)                                    & 67.41    \\
%Analytical Formula of Royer (RB) (2010)       & 67.50    \\
Modified RenB (2019)                          & 68.13    \\
MYQZR (2019)                                  & 68.22    \\
%DK1 (2009)                                    & 70.06    \\
DK2 (2018)                                    & 70.95    \\
\hline
\hline
\end{tabular}

\end{table}
The values of ROC are calculated for the step change of 0.05 MeV in $Q$-values up to 0.6 MeV. The average of all the ROC from considered formulas are listed in Table \ref{roc}, which fortifies the lesser sensitivity of the proposed formulas. This feature of our proposed formulas is indeed of great importance since $Q$-values come up with certain uncertainties from experiments, which are anticipated to impact half-life computed from various empirical formulas.\par

\begin{figure}[!htbp]
\centering
\includegraphics[width=0.8\textwidth]{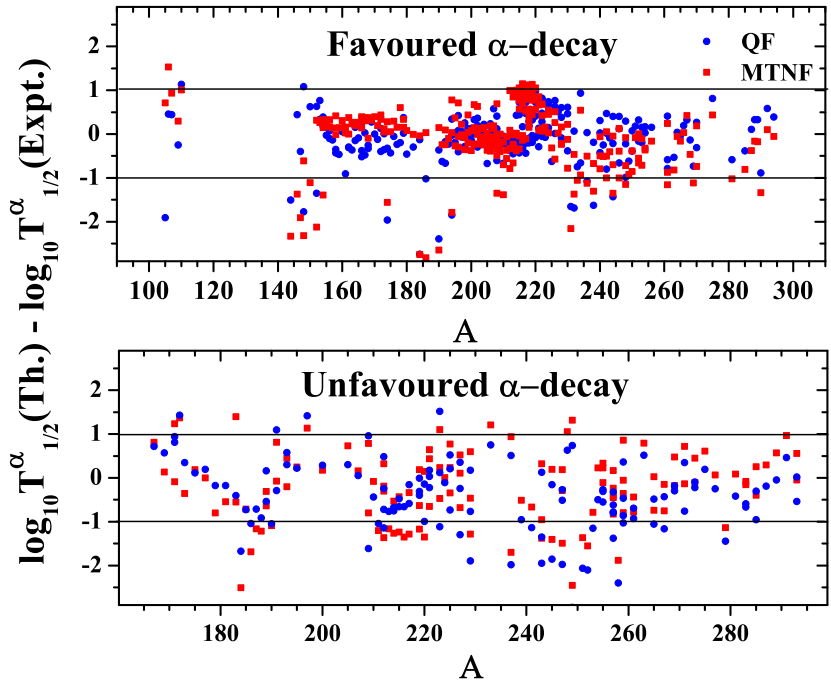}
\caption{(Colour online) Differences between logarithmic values (in sec.) of experimental data and calculated data for 267 favoured and 131 unfavoured $\alpha$-transitions calculated by using QF and MTNF formulas.}\label{fig2}
\end{figure}

Furthermore, as mentioned above, both of these formulas are fitted with the centrifugal potential term that eventually delineates favoured and unfavoured $\alpha$-decay. For an examination of QF and MTNF formulas accounting favoured and unfavoured $\alpha$-decay, incisively, we have plotted differences between logarithmic values of experimental data and calculated data in Fig. \ref{fig2} for 267 favoured and 131 unfavoured $\alpha$-transitions. The pertinency of our proposed formulas for both kind of $\alpha$-transitions is exhibited reasonably well from Fig. \ref{fig2}. \par
With the above remarks, QF and MTNF formulas, adorned with minimal coefficients, good accuracy, low uncertainty, and small sensitivity on uncertainties of $Q$-values are proficient to predict the half-lives for favoured and unfavoured $\alpha$-transitions throughout the periodic chart and also in the unknown territory of the superheavy region. On account of the above-stated aspects, in Table \ref{QF1} to Table \ref{QF15} (shown in Appendix), we have calculated $\alpha$-decay half-lives using QF formula of each possible ground to the ground-state transition of 1359 nuclei for the full range 50$\leq$Z$\leq$118 (all even and odd combination) for which experimental/estimated $Q$-values are available \cite{nndc}. To calculate $l_{min}$ for the $\alpha$-transition, spin and parities of parent and daughter nuclei are taken from latest evaluated nuclear properties table NUBASE2020 \cite{audi2020} or P. M\"{o}ller \cite{mollerparity}. The half-lives are compared with available experimental data which are found in an excellent match, and hence this comparison renders an excellent formula to the readers for a wide range of the periodic chart.\par

\begin{figure}[!htbp]
\centering
\includegraphics[width=0.8\textwidth]{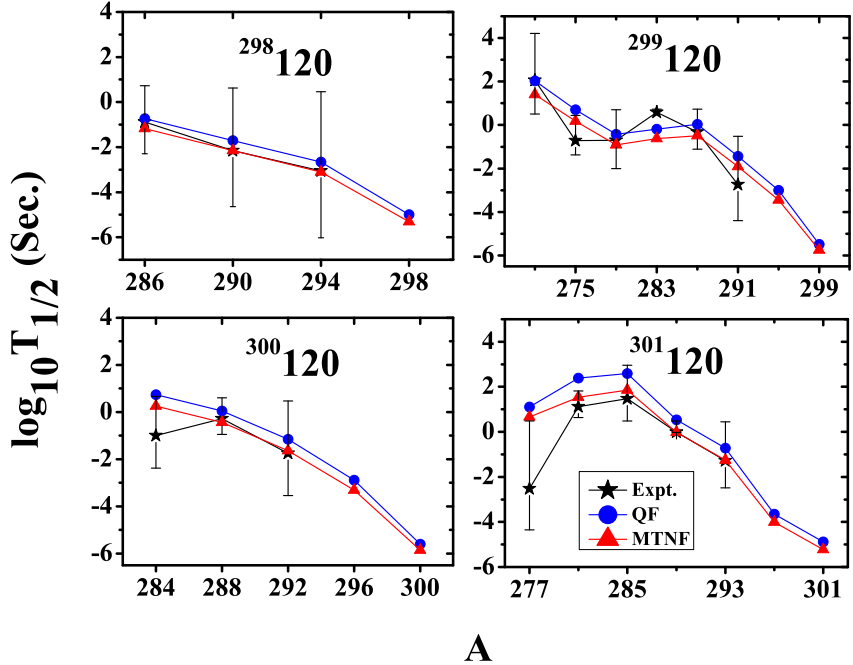}
\caption{(Colour online) $\alpha$-decay half-lives for chain of $^{298,299,300,301}$120. The experimentally data are taken from \cite{nndc}.}\label{fig3}
\end{figure}

Before applying these formulas for the prediction of the unknown superheavy region, it is desirable to judge the merit of these formulas for experimentally known  $\alpha$-decay chains. With this in view, in Fig. \ref{fig3}, we have shown half-lives for $\alpha$-decay chain of $^{298,299,300,301}$120 nuclei. These chains consist experimental decay chain of $^{294}$Og and $^{291,292,293}$Lv \cite{ogan2006,nndc,ogan2004,ogan2012}. The QF and MTNF formulas reproduce experimental data quite precisely and within the experimental uncertainty as can be ascertained from Fig. \ref{fig3}. Interestingly, on the one side, the half-life for the unknown nuclei $^{298,299,300}$120 and $^{295,296,297}$Og are found within the experimental reach and towards another side, the $\alpha$-transitions from these nuclei are found favoured, therefore reported here as potential candidates for the experiments towards an extension of the north-east corner of the periodic chart. It is worth to mentioned here that chances of production of $^{295,296,297}$Og are already pointed out by Oganessian \textit{et al.} \cite{Ogannpa2015} by the reaction of $^{48}$Ca beam and $^{251}$Cf (N = 153) target. \par

In Fig. \ref{fig3}, to calculate the half-lives from our empirical formulas the $Q$-values for unknown nuclei are taken from theoretical treatments. For accurate prediction from theoretical $Q$-values, we choose the best possible treatment among various theoretical approaches viz. RMF \cite{saxena1,saxena2,saxena3,saxena4}, FRDM \cite{moller2019}, RCHB \cite{rchb2018} and WS4 \cite{ws42014}. In addition, we use machine learning method 'XGBoost' \cite{saxenajpg2020,xgboost} to estimate $Q$-values. From these treatments, we calculate standard deviation ($\sigma$) for 1580 $Q$-values which is listed in Table \ref{table-q-alpha} in front of each treatment. \par

\begin{table}[!htbp]
\caption{\label{table-q-alpha}Standard deviation ($\sigma$) for 1580 Q$_{\alpha}$ values.}
\centering
\begin{tabular}{l@{\hskip 1.8in}c}
\hline
\hline
Theory/method&$\sigma$\\
\hline
WS4&0.3107\\
FRDM&0.4456\\
RMF&1.0097\\
XGBoost&1.7021\\
RCHB&2.014\\
\hline
\hline
\end{tabular}
\end{table}
 \begin{table}[!htbp]
 \caption{Prediction of decay chains from QF and MTNF formulas for even A isotopes of Z$=$120. The ($\ast$) values of $Q$ are taken from WS4 \cite{ws42014} mass model. (See the text for details)}
 \centering
  \resizebox{0.68\textwidth}{!}{%
%  \begin{tabular}{c|ccccc|ccc|cc}
 \begin{tabular}{c@{\hskip 0.3in}|c@{\hskip 0.3in}c@{\hskip 0.3in}c@{\hskip 0.3in}c@{\hskip 0.3in}c@{\hskip 0.3in}|c@{\hskip 0.3in}c@{\hskip 0.3in}c@{\hskip 0.3in}|c@{\hskip 0.3in}c}
 \hline
 \hline
\multicolumn{1}{c|}{Nucleus}&
 \multicolumn{5}{c|}{Expt.}&
  \multicolumn{3}{c|}{log$_{10}$T$_{1/2}$ (sec.)}&
   \multicolumn{2}{c}{Decay Modes}\\
    \cline{2-11}
    & Q& j$_{p}^{\pi}$&j$_{d}^{\pi}$& $l_{min}$ & log$_{10}$T$_{1/2}$&QF&MTNF&MBF&Predicted&Expt. \\
     &(MeV)&&&&(sec.)&($\alpha$)&($\alpha$)&(SF)&&\\
   \hline
$^{288}$120   &  13.72$^*$&  0$^{+}$ &  0$^{+}$ &  0 &           &  -6.38  &  -6.54  & 9.53 & $\alpha$ &              \\
$^{284}$Og    &  13.23$^*$&  0$^{+}$ &  0$^{+}$ &  0 &           &  -5.98  &  -6.10  & 7.25 & $\alpha$ &              \\
$^{280}$Lv    &  12.94$^*$&  0$^{+}$ &  0$^{+}$ &  0 &           &  -5.96  &  -6.00  & 5.25 & $\alpha$ &              \\
$^{276}$Fl    &  12.32$^*$&  0$^{+}$ &  0$^{+}$ &  0 &           &  -5.24  &  -5.29  & 7.07 & $\alpha$ &               \\
$^{272}$Cn    &  11.86$^*$&  0$^{+}$ &  0$^{+}$ &  0 &           &  -4.83  &  -4.86  & 5.50 & $\alpha$ &               \\
$^{268}$Ds    &  11.70    &  0$^{+}$ &  0$^{+}$ &  0 &           &  -5.05  &  -4.98  & 5.49 & $\alpha$ & $\alpha$      \\
$^{264}$Hs    &  10.59    &  0$^{+}$ &  0$^{+}$ &  0 &   -3.10   &  -3.06  &  -3.15  & 4.94 & $\alpha$ & $\alpha$/SF    \\
$^{260}$Sg    &  9.901    &  0$^{+}$ &  0$^{+}$ &  0 &   -2.31   &  -1.90  &  -2.06  & 5.74 & $\alpha$ & $\alpha$/SF    \\
\hline
$^{290}$120   &  13.70$^*$&  0$^{+}$ &  0$^{+}$ &  0 &           &  -6.34  &  -6.51  & 10.83& $\alpha$ &                \\
$^{286}$Og    &  12.92$^*$&  0$^{+}$ &  0$^{+}$ &  0 &           &  -5.37  &  -5.56  & 9.08 & $\alpha$ &                 \\
$^{282}$Lv    &  12.37$^*$&  0$^{+}$ &  0$^{+}$ &  0 &           &  -4.79  &  -4.96  & 7.05 & $\alpha$ &                   \\
$^{278}$Fl    &  12.52$^*$&  0$^{+}$ &  0$^{+}$ &  0 &           &  -5.65  &  -5.66  & 4.79 & $\alpha$ &                     \\
$^{274}$Cn    &  11.55$^*$&  0$^{+}$ &  0$^{+}$ &  0 &           &  -4.13  &  -4.24  & 10.51& $\alpha$ &                       \\
$^{270}$Ds    &  11.12    &  0$^{+}$ &  0$^{+}$ &  0 &   -4.00   &  -3.73  &  -3.81  & 7.71 & $\alpha$ & $\alpha$                \\
$^{266}$Hs    &  10.35    &  0$^{+}$ &  0$^{+}$ &  0 &   -2.64   &  -2.44  &  -2.60  & 6.11 & $\alpha$ & $\alpha$                  \\
$^{262}$Sg    &   9.60    &  0$^{+}$ &  0$^{+}$ &  0 &   -2.16   &  -1.08  &  -1.33  & 5.40 & $\alpha$ & SF                          \\
\hline
$^{292}$120   &  13.47$^*$&  0$^{+}$ &  0$^{+}$ &  0 &          &  -5.90  &  -6.12  & 12.88& $\alpha$ &            \\
$^{288}$Og    &  12.62$^*$&  0$^{+}$ &  0$^{+}$ &  0 &          &  -4.76  &  -5.01  & 10.67& $\alpha$ &             \\
$^{284}$Lv    &  11.83$^*$&  0$^{+}$ &  0$^{+}$ &  0 &          &  -3.63  &  -3.92  & 8.94 & $\alpha$ &               \\
$^{280}$Fl    &  12.23$^*$&  0$^{+}$ &  0$^{+}$ &  0 &          &  -5.04  &  -5.12  & 6.10 & $\alpha$ &                 \\
$^{276}$Cn    &  11.90    &  0$^{+}$ &  0$^{+}$ &  0 &          &  -4.91  &  -4.93  & 10.64& $\alpha$ &                   \\
$^{272}$Ds    &  10.80    &  0$^{+}$ &  0$^{+}$ &  0 &          &  -2.96  &  -3.13  & 11.43& $\alpha$ & SF                  \\
$^{268}$Hs    &   9.62    &  0$^{+}$ &  0$^{+}$ &  0 &  -0.40   &  -0.53  &  -0.86  & 9.07 & $\alpha$ & $\alpha$              \\
$^{264}$Sg    &   9.21    &  0$^{+}$ &  0$^{+}$ &  0 &  -1.43   &   0.01  &  -0.32  & 6.40 & $\alpha$ & SF                      \\
\hline
$^{294}$120   &  13.24$^*$&  0$^{+}$ &  0$^{+}$ &  0 &          &  -5.46  &  -5.72  & 15.19& $\alpha$ &        \\
$^{290}$Og    &  12.60$^*$&  0$^{+}$ &  0$^{+}$ &  0 &          &  -4.71  &  -4.97  & 12.00& $\alpha$ &         \\
$^{286}$Lv    &  11.31$^*$&  0$^{+}$ &  0$^{+}$ &  0 &          &  -2.45  &  -2.84  & 10.47& $\alpha$ &           \\
$^{282}$Fl    &  11.38$^*$&  0$^{+}$ &  0$^{+}$ &  0 &          &  -3.17  &  -3.44  & 6.55 & $\alpha$ &             \\
$^{278}$Cn    &  11.31    &  0$^{+}$ &  0$^{+}$ &  0 &          &  -3.58  &  -3.75  & 5.71 & $\alpha$ &               \\
$^{274}$Ds    &  11.70    &  0$^{+}$ &  0$^{+}$ &  0 &          &  -5.04  &  -4.98  & 12.46& $\alpha$ &                 \\
$^{270}$Hs    &   9.07    &  0$^{+}$ &  0$^{+}$ &  0 &   1.34   &   1.03  &   0.61  & 12.02& $\alpha$ & $\alpha $          \\
$^{266}$Sg    &   8.80    &  0$^{+}$ &  0$^{+}$ &  0 &   1.32   &   1.22  &   0.81  & 8.12 & $\alpha$ & SF                  \\
\hline
$^{296}$120   &  13.34$^*$&  0$^{+}$ &  0$^{+}$ &  0 &          &  -5.65  &  -5.89  & 15.10& $\alpha$ &           \\
$^{292}$Og    &  12.24$^*$&  0$^{+}$ &  0$^{+}$ &  0 &          &  -3.96  &  -4.29  & 13.64& $\alpha$ &            \\
$^{288}$Lv    &  11.29$^*$&  0$^{+}$ &  0$^{+}$ &  0 &          &  -2.40  &  -2.79  & 10.90& $\alpha$ &              \\
$^{284}$Fl    &  10.80    &  0$^{+}$ &  0$^{+}$ &  0 &  -2.60   &  -1.80  &  -2.18  & 7.63 & $\alpha$ & SF             \\
$^{280}$Cn    &  10.73    &  0$^{+}$ &  0$^{+}$ &  0 &          &  -2.20  &  -2.50  & 2.40 & $\alpha$ &                  \\
$^{276}$Ds    &  11.11    &  0$^{+}$ &  0$^{+}$ &  0 &          &  -3.70  &  -3.79  & 9.58 & $\alpha$ &                    \\
$^{272}$Hs    &   9.78    &  0$^{+}$ &  0$^{+}$ &  0 &          &  -0.95  &  -1.25  & 11.08& $\alpha$ &                      \\
$^{268}$Sg    &   8.30    &  0$^{+}$ &  0$^{+}$ &  0 &          &   2.78  &   2.30  & 10.48& $\alpha$ &                        \\
\hline
$^{298}$120   &  13.01$^*$&  0$^{+}$ &  0$^{+}$ &  0 &          &  -5.00  &   -5.31  & 12.75& $\alpha$ &  \\
$^{294}$Og    &  11.65    &  0$^{+}$ &  0$^{+}$ &  0 &          &  -2.66  &   -3.10  & 13.65& $\alpha$ & $\alpha$\\
$^{290}$Lv    &  11.00    &  0$^{+}$ &  0$^{+}$ &  0 &  -1.82   &  -1.71  &   -2.16  & 11.90& $\alpha$ & $\alpha$        \\
$^{286}$Fl    &  10.37    &  0$^{+}$ &  0$^{+}$ &  0 &  -0.80   &  -0.73  &   -1.18  &  8.66& $\alpha$ & $\alpha$/SF    \\
$^{282}$Cn    &  10.17    &  0$^{+}$ &  0$^{+}$ &  0 &  -3.30   &  -0.79  &   -1.20  &  3.24& $\alpha$ &  SF              \\
$^{278}$Ds    &  10.47    &  0$^{+}$ &  0$^{+}$ &  0 &          &  -2.14  &   -2.39  &  3.11& $\alpha$ &         \\
$^{274}$Hs    &   9.57    &  0$^{+}$ &  0$^{+}$ &  0 &          &  -0.37  &   -0.72  &  7.95& $\alpha$ &        \\
$^{270}$Sg    &   9.00    &  0$^{+}$ &  0$^{+}$ &  0 &          &   0.64  &    0.25  & 10.24& $\alpha$ &      \\
\hline
$^{300}$120   &  13.32$^*$&  0$^{+}$ &  0$^{+}$ &  0 &          &  -5.60  &   -5.86  & 12.22& $\alpha$ &             \\
$^{296}$Og    &  11.75$^*$&  0$^{+}$ &  0$^{+}$ &  0 &          &  -2.88  &   -3.31  & 13.15& $\alpha$ &             \\
$^{292}$Lv    &  10.77    &  0$^{+}$ &  0$^{+}$ &  0 &  -1.74   &  -1.15  &   -1.63  & 12.30& $\alpha$ & $\alpha$    \\
$^{288}$Fl    &  10.07    &  0$^{+}$ &  0$^{+}$ &  0 &  -0.28   &   0.05  &   -0.44  &  9.92& $\alpha$ & $\alpha$      \\
$^{284}$Cn    &   9.60    &  0$^{+}$ &  0$^{+}$ &  0 &  -1.00   &   0.74  &    0.25  &  5.00& $\alpha$ & SF            \\
$^{280}$Ds    &   9.81    &  0$^{+}$ &  0$^{+}$ &  0 &          &  -0.42  &   -0.81  &  2.24& $\alpha$ &     \\
$^{276}$Hs    &   9.28    &  0$^{+}$ &  0$^{+}$ &  0 &          &   0.44  &    0.04  &  2.54& $\alpha$ &     \\
$^{272}$Sg    &   8.70    &  0$^{+}$ &  0$^{+}$ &  0 &          &   1.54  &    1.10  &  5.75& $\alpha$ &      \\
\hline
$^{302}$120   &  12.89$^*$&  0$^{+}$ &  0$^{+}$ &  0 &          &  -4.75  &   -5.09  & 10.91& $\alpha$ &                   \\
$^{298}$Og    &  12.18$^*$&  0$^{+}$ &  0$^{+}$ &  0 &          &  -3.82  &   -4.17  & 12.68& $\alpha$ &                  \\
$^{294}$Lv    &  10.66$^*$&  0$^{+}$ &  0$^{+}$ &  0 &          &  -0.88  &   -1.38  & 13.25& $\alpha$ &                  \\
$^{290}$Fl    &   9.52$^*$&  0$^{+}$ &  0$^{+}$ &  0 &          &   1.54  &    1.00  & 11.66& $\alpha$ &                   \\
$^{286}$Cn    &   9.04$^*$&  0$^{+}$ &  0$^{+}$ &  0 &          &   2.34  &    1.81  &  6.90& $\alpha$ &                    \\
$^{282}$Ds    &   8.53$^*$&  0$^{+}$ &  0$^{+}$ &  0 &          &   3.28  &    2.76  &  3.33& $\alpha$ &            \\
$^{278}$Hs    &  8.76$^*$ &  0$^{+}$ &  0$^{+}$ &  0 &          &   1.96  &    1.48  &  0.38& SF       &            \\
$^{274}$Sg    &   8.03$^*$&  0$^{+}$ &  0$^{+}$ &  0 &          &   3.67  &    3.17  &  1.13& SF       &            \\
\hline
$^{304}$120   &  12.76$^*$&  0$^{+}$ &  0$^{+}$ &  0 &          &  -4.48  &   -4.85  &  8.70& $\alpha$ &                   \\
$^{300}$Og    &  11.96$^*$&  0$^{+}$ &  0$^{+}$ &  0 &          &  -3.34  &   -3.73  & 11.48& $\alpha$ &                   \\
$^{296}$Lv    &  10.89$^*$&  0$^{+}$ &  0$^{+}$ &  0 &          &  -1.44  &   -1.91  & 13.24& $\alpha$ &                   \\
$^{292}$Fl    &   8.95$^*$&  0$^{+}$ &  0$^{+}$ &  0 &          &   3.18  &    2.62  & 13.21& $\alpha$ &                   \\
$^{288}$Cn    &   9.11$^*$&  0$^{+}$ &  0$^{+}$ &  0 &          &   2.14  &    1.61  &  7.34& $\alpha$ &                   \\
$^{284}$Ds    &   7.88$^*$&  0$^{+}$ &  0$^{+}$ &  0 &          &   5.39  &    4.90  &  4.67&       SF &                   \\
$^{280}$Hs    &   8.02$^*$&  0$^{+}$ &  0$^{+}$ &  0 &          &   4.33  &    3.82  &  1.49&       SF &                   \\
$^{276}$Sg    &   7.60$^*$&  0$^{+}$ &  0$^{+}$ &  0 &          &   5.15  &    4.65  &  0.07&       SF &                   \\
 \hline \end{tabular}}
 \label{table-prediction-even}
  \end{table}

 \begin{table}[!htbp]
 \caption{Same as Table \ref{table-prediction-even} but for odd A isotopes of Z$=$120.}
 \centering
  \resizebox{0.85\textwidth}{!}{%
 \begin{tabular}{c@{\hskip 0.3in}|c@{\hskip 0.3in}c@{\hskip 0.3in}c@{\hskip 0.3in}c@{\hskip 0.3in}c@{\hskip 0.3in}|c@{\hskip 0.3in}c@{\hskip 0.3in}c@{\hskip 0.3in}|c@{\hskip 0.3in}c}
 \hline
 \hline
\multicolumn{1}{c|}{Nucleus}&
 \multicolumn{5}{c|}{Expt.}&
  \multicolumn{3}{c|}{log$_{10}$T$_{1/2}$ (sec.)}&
   \multicolumn{2}{c}{Decay Modes}\\
    \cline{2-11}
    & Q& j$_{p}^{\pi}$&j$_{d}^{\pi}$& $l_{min}$ & log$_{10}$T$_{1/2}$(s)&QF&MTNF&MBF&Predicted&Expt. \\
     &(MeV)&&&&(sec.)&($\alpha$)&($\alpha$)&(SF)&&\\
   \hline
\hline
$^{289}$120   &  13.71$^*$& 1/2$^{+}$& 1/2$^{+}$&  0&          &  -6.36  &  -6.52  & 12.98& $\alpha$ &               \\
$^{285}$Og    &  13.07$^*$& 1/2$^{+}$& 1/2$^{+}$&  0&          &  -5.67  &  -5.82  & 10.92& $\alpha$ &               \\
$^{281}$Lv    &  12.70$^*$& 1/2$^{+}$&13/2$^{-}$&  7&          &  -3.49  &  -4.00  & 8.52 & $\alpha$ &               \\
$^{277}$Fl    &  12.40$^*$&13/2$^{-}$& 3/2$^{+}$&  5&          &  -4.35  &  -4.60  & 10.58& $\alpha$ &               \\
$^{273}$Cn    &  11.64$^*$& 3/2$^{+}$& 1/2$^{+}$&  2&          &  -4.13  &  -4.25  & 10.43& $\alpha$ &               \\
$^{269}$Ds    &  11.51    & 1/2$^{+}$& 3/2$^{+}$&  2&  -4.42   &  -3.92  &  -4.44  & 9.20 & $\alpha$ & $\alpha$      \\
$^{265}$Hs    &  10.47    & 3/2$^{+}$& 3/2$^{+}$&  0&  -2.72   &  -2.75  &  -2.88  & 8.14 & $\alpha$ & $\alpha$      \\
$^{261}$Sg    &   9.71    & 3/2$^{+}$& 1/2$^{+}$&  2&  -0.75   &  -1.18  &  -1.44  & 7.43 & $\alpha$ & $\alpha$      \\
\hline
$^{291}$120   &  13.51$^*$& 3/2$^{+}$& 3/2$^{+}$&  0&          &  -5.98  &  -6.19  & 14.89& $\alpha$ &             \\
$^{287}$Og    &  12.80$^*$& 3/2$^{+}$& 1/2$^{+}$&  2&          &  -4.91  &  -5.17  & 12.55& $\alpha$ &             \\
$^{283}$Lv    &  12.11$^*$& 1/2$^{+}$& 5/2$^{+}$&  2&          &  -4.02  &  -4.30  & 10.83& $\alpha$ &             \\
$^{279}$Fl    &  12.43$^*$& 5/2$^{+}$&13/2$^{-}$&  5&          &  -4.41  &  -4.65  & 7.81 & $\alpha$ &             \\
$^{275}$Cn    &  11.74$^*$&13/2$^{-}$& 3/2$^{+}$&  5&          &  -3.50  &  -3.78  & 13.64& $\alpha$ &             \\
$^{271}$Ds    &  10.87    & 3/2$^{+}$& 1/2$^{+}$&  2&  -2.79   &  -2.92  &  -3.12  & 12.92& $\alpha$ & $\alpha$    \\
$^{267}$Hs    &  10.04    & 1/2$^{+}$& 3/2$^{+}$&  2&  -1.28   &  -1.43  &  -1.72  & 10.61& $\alpha$ & $\alpha$    \\
$^{263}$Sg    &   9.40    & 3/2$^{+}$& 3/2$^{+}$&  0&   0.00   &  -0.53  &  -0.82  & 7.77 & $\alpha$ & $\alpha$    \\
\hline
$^{293}$120   &  13.40$^*$& 5/2$^{+}$& 3/2$^{+}$&  2&          &  -5.55  &  -5.83  & 17.06& $\alpha$ &              \\
$^{289}$Og    &  12.59$^*$& 3/2$^{+}$& 3/2$^{+}$&  0&          &  -4.69  &  -4.95  & 14.15& $\alpha$ &              \\
$^{285}$Lv    &  11.55$^*$& 3/2$^{+}$& 1/2$^{+}$&  2&          &  -2.79  &  -3.18  & 12.19& $\alpha$ &              \\
$^{281}$Fl    &  11.82$^*$& 1/2$^{+}$& 3/2$^{+}$&  2&          &  -3.94  &  -4.16  & 8.91 & $\alpha$ &              \\
$^{277}$Cn    &  11.62    & 3/2$^{+}$&13/2$^{-}$&  5&          &  -3.22  &  -3.54  & 10.86& $\alpha$ &              \\
$^{273}$Ds    &  11.37    &13/2$^{-}$& 9/2$^{+}$&  3&  -3.77   &  -3.88  &  -3.99  & 14.33& $\alpha$ & $\alpha$     \\
$^{269}$Hs    &   9.34    & 9/2$^{+}$&11/2$^{-}$&  1&   0.99   &   0.33  &  -0.07  & 12.60& $\alpha$ & $\alpha$     \\
$^{265}$Sg    &   9.05    &11/2$^{-}$& 3/2$^{+}$&  5&   1.16   &   1.54  &   0.96  & 9.97 & $\alpha$ & $\alpha$     \\
\hline
$^{295}$120   &  13.27$^*$& 1/2$^{+}$& 5/2$^{+}$&  2&          &  -5.30  &  -5.60  & 18.35& $\alpha$ &              \\
$^{291}$Og    &  12.42$^*$& 5/2$^{+}$& 3/2$^{+}$&  2&          &  -4.12  &  -4.47  & 15.43& $\alpha$ &              \\
$^{287}$Lv    &  11.28$^*$& 3/2$^{+}$& 3/2$^{+}$&  0&          &  -2.38  &  -2.77  & 13.36& $\alpha$ &              \\
$^{283}$Fl    &  10.88$^*$& 3/2$^{+}$&11/2$^{+}$&  4&          &  -1.28  &  -1.80  & 9.71 & $\alpha$ &              \\
$^{279}$Cn    &  11.04    &11/2$^{+}$& 3/2$^{+}$&  4&          &  -2.24  &  -2.62  & 6.74 & $\alpha$ &              \\
$^{275}$Ds    &  11.40    & 3/2$^{+}$&13/2$^{-}$&  5&          &  -3.30  &  -3.54  & 13.65& $\alpha$ &              \\
$^{271}$Hs    &   9.51    &13/2$^{-}$& 3/2$^{+}$&  5&          &   0.85  &   0.27  & 14.71& $\alpha$ &              \\
$^{267}$Sg    &   8.63    & 3/2$^{+}$& 1/2$^{+}$&  2&          &   1.95  &   1.47  & 12.40& $\alpha$ &              \\
   \hline
$^{297}$120   &  13.14$^*$& 5/2$^{+}$& 1/2$^{+}$&  2&          &  -5.04  & -5.37  & 17.60& $\alpha$ &  \\
$^{293}$Og    &  12.24$^*$& 1/2$^{+}$& 5/2$^{+}$&  2&          &  -3.74  & -4.12  & 17.39& $\alpha$ & \\
$^{289}$Lv    &  11.10    & 5/2$^{+}$& 3/2$^{+}$&  2&          &  -1.74  & -2.21  & 14.14& $\alpha$ &  \\
$^{285}$Fl    &  10.56    & 3/2$^{+}$& 3/2$^{+}$&  0&  -0.82   &  -1.21  & -1.63  & 10.98& $\alpha$ & $\alpha$ \\
$^{281}$Cn    &  10.45    & 3/2$^{+}$& 9/2$^{+}$&  4&  -0.89   &  -0.80  & -1.30  &  4.86& $\alpha$ & $\alpha$ \\
$^{277}$Ds    &  10.83    & 9/2$^{+}$& 3/2$^{+}$&  4&  -2.39   &  -2.32  & -2.63  &  9.40& $\alpha$ & $\alpha$\\
$^{273}$Hs    &   9.70    & 3/2$^{+}$&13/2$^{-}$&  5&  -0.12   &   0.33  & -0.21  & 12.18& $\alpha$ & $\alpha$\\
$^{269}$Sg    &   8.65    &13/2$^{-}$& 9/2$^{+}$&  3&   2.27   &   2.11  &  1.58  & 12.85& $\alpha$ & $\alpha$\\
\hline
$^{299}$120   &  13.26$^*$& 1/2$^{+}$& 1/2$^{+}$&  0&          &  -5.49  & -5.75  & 15.41& $\alpha$ &              \\
$^{295}$Og    &  11.90$^*$& 1/2$^{+}$& 5/2$^{+}$&  2&          &  -3.00  & -3.45  & 15.95& $\alpha$ &              \\
$^{291}$Lv    &  10.89    & 5/2$^{+}$& 5/2$^{+}$&  0&  -2.20   &  -1.45  & -1.91  & 15.09& $\alpha$ & $\alpha$       \\
$^{287}$Fl    &  10.16    & 5/2$^{+}$& 3/2$^{+}$&  2&  -0.29   &   0.03  & -0.50  & 12.14& $\alpha$ & $\alpha$      \\
$^{283}$Cn    &   9.94    & 3/2$^{+}$& 3/2$^{+}$&  0&   0.60   &  -0.19  & -0.63  &  6.94& $\alpha$ & $\alpha$/SF   \\
$^{279}$Ds    &  10.08    & 3/2$^{+}$& 9/2$^{+}$&  4&  -0.74   &  -0.43  & -0.92  &  4.39& $\alpha$ & SF      \\
$^{275}$Hs    &   9.44    & 9/2$^{+}$& 3/2$^{+}$&  4&  -0.82   &   0.70  &  0.17  &  9.04& $\alpha$ & $\alpha$\\
$^{271}$Sg    &   8.89    & 3/2$^{+}$&13/2$^{-}$&  5&   2.16   &   2.03  &  1.40  & 11.24& $\alpha$ & $\alpha$    \\
\hline
$^{301}$120   &  13.06$^*$& 3/2$^{+}$& 1/2$^{+}$&  2&          &  -4.88  & -5.23 & 14.58& $\alpha$ &                   \\
$^{297}$Og    &  12.10$^*$& 1/2$^{+}$& 1/2$^{+}$&  0&          &  -3.65  & -4.01 & 15.81& $\alpha$ &                   \\
$^{293}$Lv    &  10.68    & 1/2$^{+}$& 3/2$^{+}$&  2&  -1.28   &  -0.72  & -1.26 & 15.59& $\alpha$ & $\alpha$           \\
$^{289}$Fl    &   9.97    & 3/2$^{+}$& 5/2$^{+}$&  2&  -0.01   &  0.53  & -0.02 & 13.50& $\alpha$ & $\alpha$           \\
$^{285}$Cn    &   9.32    & 5/2$^{+}$&15/2$^{-}$&  5&   1.48   &  2.59  & 1.85 &  9.05& $\alpha$ & $\alpha$            \\
$^{281}$Ds    &   9.51    &15/2$^{-}$& 3/2$^{+}$&  7&   1.11   &  2.38  & 1.53 &  5.67& $\alpha$ &                  \\
$^{277}$Hs    &   9.05    & 3/2$^{+}$& 3/2$^{+}$&  0&  -2.52   &  1.11  & 0.67 &  4.81& $\alpha$ & $\alpha$    \\
$^{273}$Sg    &   8.20$^*$& 3/2$^{+}$& 9/2$^{+}$&  4&          &  3.81  & 3.17 &  6.99& $\alpha$ &                  \\
\hline
$^{303}$120   &  12.81$^*$& 1/2$^{+}$& 3/2$^{+}$&  2&          &  -4.37 & -4.78  & 12.88& $\alpha$ &                   \\
$^{299}$Og    &  12.05$^*$& 3/2$^{+}$& 1/2$^{+}$&  2&          &  -3.32 & -3.75  & 15.06& $\alpha$ &                   \\
$^{295}$Lv    &  10.77$^*$& 1/2$^{+}$& 7/2$^{+}$&  4&          &  -0.44 & -1.07  & 16.36& $\alpha$ &                   \\
$^{291}$Fl    &   9.27$^*$& 7/2$^{+}$& 1/2$^{+}$&  4&          &  2.96 & 2.26  & 15.02& $\alpha$ &                   \\
$^{287}$Cn    &   9.06$^*$& 1/2$^{+}$& 5/2$^{+}$&  2&          &  2.48 & 1.90  & 10.89& $\alpha$ &                   \\
$^{283}$Ds    &   8.16$^*$& 5/2$^{+}$&15/2$^{-}$&  5&          &  5.53 & 4.79  &  6.88& $\alpha$ &                   \\
$^{279}$Hs    &   8.38$^*$&15/2$^{-}$& 3/2$^{+}$&  7&          &  5.13 & 4.21  &  3.83& $\alpha$ &           \\
$^{275}$Sg    &   7.86$^*$& 3/2$^{+}$&11/2$^{+}$&  4&          &  4.96 & 4.31  &  4.16& $\alpha$ &                \\
 \hline
  \end{tabular}}
   \label{table-prediction-odd}
  \end{table}

This table itself depicts the exactitude of WS4 \cite{ws42014} mass model over the above-mentioned theories for the range 50$\leq$Z$\leq$118, henceforth, we use QF and MTNF formulas for the precise prediction of decay chains of Z$=$120 isotopes in Tables \ref{table-prediction-even} and \ref{table-prediction-odd} using $Q$-values from WS4 mass model (if experimental values are not available). As mentioned above, to explore the competition between $\alpha$-decay and spontaneous fission half-lives, we have also evaluated half-lives for spontaneous fission from modified Bao formula (MBF) \cite{saxenajpg2020} (Eqn. \ref{baoSF}). The probable decay-modes of Z$=$120 isotopes with 168$\leq$N$\leq$184 and the corresponding half-lives are mentioned in Tables \ref{table-prediction-even} and \ref{table-prediction-odd} along with available experimental half-lives and decay modes \cite{nndc}. The ($\ast$) values of $Q$ are taken from WS4 \cite{ws42014} mass model. To calculate $l_{min}$ for the $\alpha$-transition (from Eqn. \ref{lmin}), spin and parities of parent and daughter nuclei are taken from latest evaluated nuclear properties table NUBASE2020 \cite{audi2020} or P. M\"{o}ller \cite{mollerparity}. Remarkably, our predictions are in an excellent match with available experimental data.\par
\section{Conclusions}
A new empirical formula (QF) in conjunction with one modified formula (MTNF) are reported in this article to describe both favoured and unfavoured $\alpha$-transitions. Experimental data of a total of 398 nuclei have been used for fitting and subsequently, the formulas are probed on various crucial aspects required for estimating $\alpha$-decay half-lives. The proposed formulas are also established on the ground of accuracy, uncertainty, least no. of coefficients, less ineffectual on the uncertainties observed in the $Q$-values, and efficacious for all odd-even combination of Z-N throughout the periodic chart. As a utilization, half-lives of various unknown favoured and unfavoured $\alpha$-transitions are predicted along with probable decay chains of Z$=$120 isotopes. Our predictions are found in an excellent match with available experimental data and anticipated as more precise ingratiation for the future experiments towards synthesising new superheavy elements and isotopes.
\section{Acknowledgement}
The support provided by SERB (DST), Govt. of India under CRG/2019/001851 is acknowledged. Authors are indebted to Prafulla Saxena regarding the various discussion on machine learning.

\appendix
\section{}
\begin{small}
We have calculated $\alpha$-decay half-lives using QF formula of each possible ground to ground state transition of 1359 nuclei for the full range 50$\leq$Z$\leq$118 (all even and odd combination) for which experimental/estimated $Q$-values (in MeV) are available \cite{nndc}. Table \ref{QF1} to Table \ref{QF8} consist of favoured $\alpha$-decay transitions whereas Table \ref{QF9} to Table \ref{QF15} represents unfavoured $\alpha$-decay transitions. To calculate $l_{min}$ for the $\alpha$-transition [Using equation \ref{lmin}], spin and parities of parent and daughter nuclei are taken from latest evaluated nuclear properties table NUBASE2020 \cite{audi2020} or P. M\"{o}ller \cite{mollerparity}. The half-lives are compared with available experimental data \cite{nndc}.
\end{small}
%%%%%%%%%%%%%%%%%%%%%%%%%%%%%%%%%%%%%%%%%%%%%%%%%%%%%%%%%%%%%%%%%%%%%%%%%%%%%%%%%%%%%%%%%%%%%%%%%%%%%%%%%%%%%%%%%%%%%%%%%%%%%%%%%%%%%%%%%%%%%%%%%%%%%%%%%%
%%%%%%%%%%%%%%%%%%%%%%%%%%%%%%%%%%%%%%%%%%%%%%%%%%%%%%%%%%%%%%%%%%%%%%%%%%%%%%%%%%%%%%%%%%%%%%%%%%%%%%%%%%%%%%%%%%%%%%%%%%%%%%%%%%%%%%%%%%%%%%%%%%%%%
\begin{table}[!htbp]
\caption{$\alpha$-decay half-lives (in sec.) from QF formula for ground to ground favoured transitions in the range 50$\leq$Z$\leq$118.}
\centering
\def\arraystretch{1.15}
\resizebox{1.00\textwidth}{!}{%
%{
}
}
\label{QF1}
\end{table}
%%%%%%%%%%%%%%%%%%%%%%%%%%%%%%%%%%%%%%%%%%%%%%%%%%%%%%%%%%%%%%%%%%%%%%%%%%%%%%%%%%%%%%%%%%%%%%%%%%%%%%%%%%%%%%%%%%%%%%%%%%%%%%%%%%%%%%%%%%%%%%%%%
\begin{table}[!htbp]
\caption{Continue from Table \ref{QF1}.}
\centering
\def\arraystretch{1.15}
\resizebox{1.00\textwidth}{!}{%
%{%
}
}
\label{QF2}
\end{table}
%%%%%%%%%%%%%%%%%%%%%%%%%%%%%%%%%%%%%%%%%%%%%%%%%%%%%%%%%%%%%%%%%%%%%%%%%%%%%%%%%%%%%%%%%%%%%%%%%%%%%%%%%%%%%%%%%%%%%%%%%%%%%%%%%%%%%%%%%%%%%%%%%%%%%%%%%%%
\begin{table}[!htbp]
\caption{Continue from Table \ref{QF2}.}
\centering
\def\arraystretch{1.15}
\resizebox{1.00\textwidth}{!}{%
%{%
}
}
\label{QF3}
\end{table}
%%%%%%%%%%%%%%%%%%%%%%%%%%%%%%%%%%%%%%%%%%%%%%%%%%%%%%%%%%%%%%%%%%%%%%%%%%%%%%%%%%%%%%%%%%%%%%%%%%%%%%%%%%%%%%%%%%%%%%%%%%%%%%%%%%%%%%%%%%%%%%%%%%%%%%%%%%%%%%%%%%%%%%%%%%%%%%
\begin{table}[!htbp]
\caption{Continue from Table \ref{QF3}.}
\centering
\def\arraystretch{1.15}
\resizebox{1.00\textwidth}{!}{%
%{%
}
}
\label{QF4}
\end{table}
%%%%%%%%%%%%%%%%%%%%%%%%%%%%%%%%%%%%%%%%%%%%%%%%%%%%%%%%%%%%%%%%%%%%%%%%%%%%%%%%%%%%%%%%%%%%%%%%%%%%%%%%%%%%%%%%%%%%%%%%%%%%%%%%%%%%%%%%%%%%%%%%%%%%%%%%%%
\begin{table}[!htbp]
\caption{Continue from Table \ref{QF4}.}
\centering
\def\arraystretch{1.15}
\resizebox{1.00\textwidth}{!}{%
%{%
}
}
\label{QF5}
\end{table}
%%%%%%%%%%%%%%%%%%%%%%%%%%%%%%%%%%%%%%%%%%%%%%%%%%%%%%%%%%%%%%%%%%%%%%%%%%%%%%%%%%%%%%%%%%%%%%%%%%%%%%%%%%%%%%%%%%%%%%%%%%%%%%%%%%%%%%%%%%%%%%%%%%%%%
\begin{table}[!htbp]
\caption{Continue from Table \ref{QF5}.}
\centering
\def\arraystretch{1.15}
\resizebox{1.00\textwidth}{!}{%
%{%
}
}
\label{QF6}
\end{table}
%%%%%%%%%%%%%%%%%%%%%%%%%%%%%%%%%%%%%%%%%%%%%%%%%%%%%%%%%%%%%%%%%%%%%%%%%%%%%%%%%%%%%%%%%%%%%%%%%%%%%%%%%%%%%%%%%%%%%%%%%%%%%%%%%%%%%%%%%%%%%%%%%%%%%%%%
\begin{table}[!htbp]
\caption{Continue from Table \ref{QF6}.}
\centering
\def\arraystretch{1.15}
\resizebox{1.00\textwidth}{!}{%
%{%
}
}
\label{QF7}
\end{table}
%%%%%%%%%%%%%%%%%%%%%%%%%%%%%%%%%%%%%%%%%%%%%%%%%%%%%%%%%%%%%%%%%%%%%%%%%%%%%%%%%%%%%%%%%%%%%%%%%%%%%%%%%%%%%%%%%%%%%%%%%%%%%%%%%%%%%%%%%%%%%%%%%%%%%%%%
\begin{table}[!htbp]
\caption{Continue from Table \ref{QF7}.}
\centering
\def\arraystretch{1.15}
\resizebox{1.00\textwidth}{!}{%
%{%
}
}
\label{QF8}
\end{table}

%%%%%%%%%%%%%%%%%%%%%%%%%%%%%%%%%%%%%%%%%%%%%%%%%%%%%%%%%%%%%%%%%%%%%%%%%%%%%%%%%%%%%%%%%%%%%%%%%%%%%%%%%%%%%%%%%%%%%%%%%%%%%%%%%%%%%%%%%%%%%%%%%%%%%%%%
\begin{table}[!htbp]
\caption{$\alpha$-decay half-lives (in sec.) from QF formula for ground to ground unfavoured transitions in the range 50$\leq$Z$\leq$118.}
\centering
\def\arraystretch{1.15}
\resizebox{1.00\textwidth}{!}{%
%{%
} }
\label{QF9}
\end{table}
%%%%%%%%%%%%%%%%%%%%%%%%%%%%%%%%%%%%%%%%%%%%%%%%%%%%%%%%%%%%%%%%%%%%%%%%%%%%%%%%%%%%%%%%%%%%%%%%%%%%%%%%%%%%%%%%%%%%%%%%%%%%%%%%%%%%%%%%%%%%%%%%%%%%
\begin{table}[!htbp]
\caption{Continue from Table \ref{QF9}.}
\centering
\def\arraystretch{1.15}
\resizebox{1.00\textwidth}{!}{%
%{%
} }
\label{QF10}
\end{table}
%%%%%%%%%%%%%%%%%%%%%%%%%%%%%%%%%%%%%%%%%%%%%%%%%%%%%%%%%%%%%%%%%%%%%%%%%%%%%%%%%%%%%%%%%%%%%%%%%%%%%%%%%%%%%%%%%%%%%%%%%%%%%%%%%%%%%%%%%%%%%%%%%%%%%%
\begin{table}[!htbp]
\caption{Continue from Table \ref{QF10}.}
\centering
\def\arraystretch{1.15}
\resizebox{1.00\textwidth}{!}{%
%{%
} }
\label{QF11}
\end{table}
%%%%%%%%%%%%%%%%%%%%%%%%%%%%%%%%%%%%%%%%%%%%%%%%%%%%%%%%%%%%%%%%%%%%%%%%%%%%%%%%%%%%%%%%%%%%%%%%%%%%%%%%%%%%%%%%%%%%%%%%%%%%%%%%%%%%%%%%%%%%%%%%%%%%%%%%%
\begin{table}[!htbp]
\caption{Continue from Table \ref{QF11}.}
\centering
\def\arraystretch{1.15}
\resizebox{1.00\textwidth}{!}{%
%{%
} }
\label{QF12}
\end{table}
%%%%%%%%%%%%%%%%%%%%%%%%%%%%%%%%%%%%%%%%%%%%%%%%%%%%%%%%%%%%%%%%%%%%%%%%%%%%%%%%%%%%%%%%%%%%%%%%%%%%%%%%%%%%%%%%%%%%%%%%%%%%%%%%%%%%%%%%%%%%%%%%%%%%%%%
\begin{table}[!htbp]
\caption{Continue from Table \ref{QF12}.}
\centering
\def\arraystretch{1.15}
\resizebox{1.00\textwidth}{!}{%
%{%
} }
\label{QF13}
\end{table}
%%%%%%%%%%%%%%%%%%%%%%%%%%%%%%%%%%%%%%%%%%%%%%%%%%%%%%%%%%%%%%%%%%%%%%%%%%%%%%%%%%%%%%%%%%%%%%%%%%%%%%%%%%%%%%%%%%%%%%%%%%%%%%%%%%%%%%%%%%%%%%%%%%%%%%%%
\begin{table}[!htbp]
\caption{Continue from Table \ref{QF13}.}
\centering
\def\arraystretch{1.15}
\resizebox{1.00\textwidth}{!}{%
%{%
} }
\label{QF14}
\end{table}
%%%%%%%%%%%%%%%%%%%%%%%%%%%%%%%%%%%%%%%%%%%%%%%%%%%%%%%%%%%%%%%%%%%%%%%%%%%%%%%%%%%%%%%%%%%%%%%%%%%%%%%%%%%%%%%%%%%%%%%%%%%%%%%%%%%%%%%%%%%%%%%%%%%%%%%%
\begin{table}[!htbp]
\caption{Continue from Table \ref{QF14}.}
\centering
\def\arraystretch{1.15}
\resizebox{1.00\textwidth}{!}{%
%{%
} }
\label{QF15}
\end{table}

\end{document}